\documentclass[12pt]{iopart}
\usepackage{amssymb,graphicx}

\renewcommand{\vec}{\mathbf}

\begin{document}

\title[Success of T-matrix Theories for a Modified Hubbard Model]{Analyzing the success of T-matrix diagrammatic theories in representing a modified Hubbard model}

\author{P. Pisarski and R.J. Gooding}
\address{Department of Physics, Engineering Physics, and Astronomy,\\ 
Queen's University, Kingston, ON K7L 3N6, Canada\ead{pawel.pisarski@gmail.com}}

\date{\today}

\begin{abstract}
We present a systematic study of various forms of renormalization
that can be applied in the calculation of the self-energy of the Hubbard
model within the T-matrix approximation. We compare the exact solutions 
of the attractive and repulsive Hubbard models, for linear chains of lengths up to
eight sites, with all possible taxonomies of the T-matrix approximation. 
For the attractive Hubbard model, the success of a minimally self-consistent theory
found earlier in the atomic limit (Phys. Rev. B {\bf 71}, 155111 (2005)) 
is not maintained for finite clusters unless one is in the very strong correlation limit.
For the repulsive model, in the weak correlation limit at low electronic densities -- that is, 
where one would expect a self-consistent T-matrix theory
to be adequate -- we find the fully renormalized theory to be most successful.
In our studies we employ a modified Hubbard interaction that eliminates all Hartree 
diagrams, an idea which was proposed earlier (Phys. Rev. B {\bf 63}, 035104 (2000)).
\end{abstract}


\maketitle

\section{ \label{SEC:introduction} Introduction }

The Hubbard model was originally proposed to describe interacting electrons in solids, and the transition 
between conducting and insulating systems \cite{hubbard,hubbard:2}. The early work of Kanamori \cite{kanamori}
examined the instability towards ferromagnetic ordering in narrow band systems, and work continues
using this model in investigations of such systems \cite{hlubina}.
It has been extensively studied for many years, and has been a focus of interest in the theory of 
high-temperature superconductivity \cite{PWA,micnas}, 
and more recently for ultra-cold atoms trapped in optical lattices \cite{fisher,jaksch}.
Even though the Hubbard Hamiltonian seems to be a minimal model examining interacting quantum systems, its exact solution can be given only in a few cases.
Therefore, many approximate theories have been developed to describe the thermodynamic and dynamical
properties of this model.

One of the best known and broadly used approximate methods in condensed matter (and nuclear) physics is the 
T-matrix approximation \cite{FW}.
One way to arrive at this approximation is to derive Dyson's equation within the framework of the equation of motion method.
This method results in varying levels of self-consistency -- some propagators are renormalized in a self-consistent fashion
whereas others are not.  The pioneering work of Kadanoff and Baym \cite{kadanoff:1961, baym:1961, baym:1962} has given rise to the question of what level of self-consistency is appropriate.
The purpose of our work is to present a systematic study of these various approximation schemes that can be applied in the calculation of the self-energy within the T-matrix approximation.
To be concrete,  we   compare the exact solution of the model Hamiltonian with all possible taxonomies (see below) that emerge from the T-matrix approximation.
This will allow us to review candidate theories and decide which one reproduces the exact results best. Our work closely
follows that of an earlier paper \cite{verga} that focused on such comparisons for the case of the attractive Hubbard model in the atomic limit.
Here we extend such work to systems of finite extent, specifically linear chains, and also consider the case of the repulsive Hubbard model. 
First, we briefly review each of these models.

The repulsive Hubbard model (RHM) has been studied with reference to the Mott metal-insulator transition \cite{hubbard,hubbard:2,imadaRMP} and such transitions remain a great theoretical challenge.
The model may also lead to a variety of interesting magnetic correlations, 
such as ferromagnetism \cite{kanamori,hlubina} or antiferromagnetism \cite{Igoshev2010}.
It may be able to describe other phenomena, for example colossal magneto-resistance transitions \cite{imadaRMP,CMR}.
Many authors have used the T-matrix approximation to investigate the properties of the RHM, including extensive work which has been completed within the non-self-consistent approximation \cite{fukuyama:1, fukuyama:2}. 
Also, as mentioned above, and as first proposed by Anderson \cite{PWA}, this model is believed to be the simplest model that
may account for the pairing instability in high-temperature superconductors \cite{bickers:1,bickers:2,kobayashi,yanase}.
 
Changing the sign of the on-site interaction leads to the attractive Hubbard model (AHM), and this
model has been extensively studied in the theory of superconductivity.
A wide variety of superconducting phenomena have been successfully explained, owing to the theory of Bardeen, Cooper and Schrieffer (BCS) \cite{bcs}.
Within that theory, superconductivity arises upon formation of an order parameter \cite{ginzburg}.
Some authors \cite{alexandrov, feinberg, emin, ranninger} suggested that electrons can form pairs already above $T_c$,
but superconductivity is not the stable thermodynamic phase because of large phase fluctuations of the order parameter.
According to such ideas, the transition to the superconducting state appears at a temperature at which the pairs acquire a common phase \cite{emery}. Therefore, one approach is to focus on the pairing fluctuations above $T_c$, as they may lead to a non-Fermi liquid like behaviour \cite{levin}. It has been suggested that due to the pairing fluctuations in normal state one can observe anomalous properties, like the pseudo-gap, in high $T_c$ materials \cite{levin}.
We mention the above theories because, to understand the physics behind the problem of paring fluctuations, one can 
employ the T-matrix approximation to the AHM.
Some authors \cite{fresard, haussmann,haussmann:2} examined the problem using one variant of the taxonomy of
the T-matrix approximation, namely when all propagators are fully renormalized (referred to as the fully
self-consistanet T-matrix theory).
This might be expected to be the most accurate case, quite simply because this approximation contains the most diagrams, and therefore should result in the most accurate approximation. In contrast to this,
based on earlier works of Patton and others \cite{patton,patton:2, schmid, marcelja},
Levin and coworkers \cite{levin} developed a theory utilizing the idea that vertex corrections omitted in the T-matrix approximation can cancel some of the contributions to the self-energy that are retained in the fully self-consistent case.
That has lead them to suggest a specific combination of single particle propagators, both bare and fully renormalized (another variant in the taxonomy of T-matrix theories) in the equations for the pair susceptibility and the self energy.
This combination is consistent with some ideas expressed by Kadanoff and Baym \cite{kadanoff:1961}.
Earlier work by one of us and colleagues \cite{verga}, which has been done only for the atomic limit (which can be thought of as working in the infinite strong coupling limit),
provided numerical support for the use of the so-called minimally self-consistent theories, in agreement
with this less-is-better approach \cite{kadanoff:1961}.

Encouraged by the results of that work \cite{verga}, here we examine the problem of the AHM once again, but this time for all coupling regimes (from weak to strong coupling).
Quite simply, this means going beyond the atomic limit and examining when the interactions are weak, intermediate or strong \textit{relative} to the non-interacting bandwidth, the latter giving
a quantitative estimate of the scale of the kinetic energy.
Other comparisons of approximate theories applied to the Hubbard model exist, including
(i) numerical studies within the dynamical mean-field theory \cite{potthoff}, and (ii) the analysis of the half-filled Hubbard model in strong coupling regime and its transformation into the $t$-$J$ model \cite{gingras}. Other work has been completed that, again, compares results from small clusters to T-matrix theories
\cite{verdozzi1}, and very recently work has been completed on self-consistent T-matrix theories by studying double occupancies in small clusters 
\cite{verdozzi2}, in particular in relation to the Kadanoff-Baym equations.
Our paper contains work in a similar  vein -- comparing exact solutions of small systems to approximate theories of models of strongly correlated electrons moving on lattices.

Our paper is organized as follows.
In the next section we introduce the model Hamiltonian, as well as a modified Hubbard model that we use. In \S \ref{SEC:tmatrix} we review the T-matrix approximation, and introduce the various schemes (members of the
taxonomy) for evaluating the self-energy within this approximation.
In \S \ref{SEC:RHM} and \S \ref{SEC:AHM} we present comparisons of the results obtained from the exact diagonalization of the Hamiltonian matrix and the T-matrix approximation.
Finally, \S \ref{SEC:conclusions} contains a summary of our results and conclusions.

\section{ \label{SEC:hamiltonian} Modified Hubbard Hamiltonian }

Our work focuses on comparisons of the predictions of diagrammatic T-matrix calculations with exact results for small clusters.
We study one dimensional chains described by the Hubbard Hamiltonian
\begin{equation}
\hat{\mathcal{H}}=
 -t \sum_{\langle i,j \rangle \sigma} \left( \hat{c}^\dagger_{i\sigma} \hat{c}_{j\sigma} + H.c. \right)
 +U\sum_i \hat{n}_{i\uparrow} \hat{n}_{i\downarrow}.
\label{EQ:hamiltonian_Hubbard}
\end{equation}
We work in the grand canonical ensemble in which the thermal average of the electron density is determined by the chemical potential. The
grand Hamiltonian is then given by
\begin{equation}
\hat{\mathcal{K}}~=~
\hat{\mathcal{H}}
~-~\mu \sum_{i,\sigma} \hat{n}_{i\sigma}  
\label{EQ:Khamiltonian_Hubbard}
\end{equation}
In (\ref{EQ:hamiltonian_Hubbard}) the first term describes the hopping of an electron between the neighbouring sites, where $t$ is the single particle hopping integral and
$\hat{c}^\dagger_{i\sigma}$, $\hat{c}_{i\sigma}$ create and annihilate, respectively, a fermion of spin $\sigma$ on site $i$;
the second term describes the interaction, of strength $U$, between electrons residing on the same site --
if the interaction is repulsive it costs an energy $U>0$ for two electrons to remain on the same site
and if it is attractive the energy of the system is lowered by $U<0$ if two electrons occupy the same site. 
The second term in (\ref{EQ:Khamiltonian_Hubbard}) involves the chemical potential, $\mu$, with $\hat{n}_{i\sigma}=\hat{c}^\dagger_{i\sigma} \hat{c}_{i\sigma}$ representing the spin-resolved number operator.
 
In order to complete a comparison of oft-used T-matrix many-body theories to the Hubbard model one is faced with the dilemma of the vast number of diagrams that must be included,
even in low-order expansions of the self energy, \textit{if} the Hartree bubble is included.
A purely theoretical modification, which should not be associated with any physical system, was proposed by Zlati\'{c}~\emph{et~al.} \cite{zlatic}.
These authors begin with (what we will call) the Zlatic Hamiltonian, being written in the form
\begin{equation}\label{eq:ZlHam}
\hat{\mathcal{H}}_{Zl}=
 -t \sum_{\langle i,j \rangle \sigma} \left( \hat{c}^\dagger_{i\sigma} \hat{c}_{j\sigma} + H.c. \right)
 +U\sum_i \left( \hat{n}_{i\uparrow}-\frac{n}{2} \right) \left( \hat{n}_{i\downarrow}-\frac{n}{2} \right).
\end{equation}
and associated grand Hamiltonian
\begin{equation}
\hat{\mathcal{K}}_{Zl}~=~
\hat{\mathcal{H}}_{Zl}
~-~\mu' \sum_{i,\sigma} \hat{n}_{i\sigma}  
\label{eq:ZlGrandHam}
\end{equation}
The idea of using these forms comes from the realization that the sum over all one-legged (Hartree) diagrams for the self energy equals zero.
Put another way, subtracting the mean-field expectation value of the electron density of a given spin species in the paramagnetic state 
-- that is, $\langle \hat{n}_{i\sigma} \rangle = n_{\sigma} = n/2$ -- from the electron-density operator in the 
interaction term makes it possible to avoid all Hartree diagrams. A relation between the original and the Zlatic grand
Hamiltonians can be obtained by noticing their similarity when 
\begin{equation}\label{eq:ZlaticTran}
\mu'=\mu-U \frac{n}{2},
\end{equation}
That is, with this substitution in (\ref{eq:ZlGrandHam}), apart from a constant term, one recovers the usual Hubbard grand Hamiltonian.
This modified Hamiltonian has also been used in an earlier publication, as mentioned in the introduction,
on a related study of T-matrix approximations in the atomic limit \cite{verga}.

It is apparent that the Zlatic Hamiltonian of (\ref{eq:ZlHam}) could have non-standard thermodynamic
properties. That is, the Hamiltonian includes an interaction term which is dependent on the local electron density,
which itself has to that have to be determined self consistently from
taking a thermal expectation value of the electron density operator. 
Therefore, unlike the original and familiar Hubbard interaction in (\ref{EQ:hamiltonian_Hubbard}),
or the form used in Monte Carlo studies \cite{sedgewick} for which particle-hole symmetry is
an added benefit (for such studies, $\hat{n}_{i\sigma}-n/2$ is replaced by $\hat{n}_{i\sigma}-1/2$ in (\ref{EQ:hamiltonian_Hubbard})),
in the model of (\ref{eq:ZlHam}) the interaction is modified by the self-consistently determined electron density.
(Note that we are assuming spatially uniform and paramagnetic phases).
As we show below, this leads to unusual results when the density is calculated as a function of $\mu'$.
Specifically, we show that when one works with this unusual Hamiltonian one obtains an unphysical first-order phase transition at low temperatures.
 
\section{ \label{SEC:tmatrix} T-Matrix Approximation }

Some background has been discussed in the introduction, so here we simply define the
equations corresponding to the various T-matrix approximations that we use in our calculations.
(Also, see \cite{verga} for related details of such work.)  
Of course, one has to remember that this theory is expected to be valid only
when it describes the effects of the effective interaction when the electronic density is either close to zero ($n\approx 0$, $n$ being the average electronic density per site), or close to complete filling ($n\approx 2$) \cite{FW}.

For the model discussed in (\ref{eq:ZlHam}), the T-matrix self-energy is given by
\begin{equation} \label{EQ:sigma}
\Sigma(\vec{k}, i \omega_m) = -U^2 \frac{T}{L} \sum_{\vec{q}, l} \frac{\chi_0(\vec{q}, i \nu_l)}{1+U \chi_0(\vec{q}, i \nu_l)} G_c(\vec{q}-\vec{k}, i \nu_l - i \omega_m),
\end{equation}
where $T$ is the temperature, $L$ is the number of lattice sites, and $i \omega_m=i \pi T (2m+1)$ and $i \nu_l=i 2\pi T l$ are Fermion and Boson Matsubara's frequencies, respectively.
The bare pair susceptibility $\chi_0(\vec{q},i \nu_l)$ is a convolution of two single-particle propagators, and is given by
\begin{equation} \label{EQ:chi0}
\chi_0(\vec{q},i \nu_l) = \frac{T}{L} \sum_{\vec{k}, m} G_a(\vec{k},i \omega_m) G_b(\vec{q}-\vec{k},i \nu_l - i \omega_m).
\end{equation}
It is important to note that we have introduced subscripts -- $a$, $b$ and $c$ -- to differentiate between what
we will call the various taxonomies of the T-matrix approximation \cite{verga}.
These subscripts can be either ``$0$", to indicate that the non-interacting Green's function is used, \emph{viz.}
\begin{equation} \label{EQ:green0}
G_0(\vec{k}, i \omega_m) = \frac{1}{i \omega_m -\epsilon_{\vec{k}} + \mu'},
\end{equation}
or the subscript can be absent, which is used to indicate that the fully renormalized Green's function is used, \emph{viz.}
\begin{equation} \label{EQ:green}
G(\vec{k}, i \omega_m) = \frac{1}{i \omega_m -\epsilon_{\vec{k}} + \mu' - \Sigma(\vec{k}, i \omega_m)}.
\end{equation}

We have examined all possible variations of $a,~b$ and $c$ in (\ref{EQ:sigma},\ref{EQ:chi0}), which we will refer to
as all possible taxonomies.
To differentiate between those six possible T-matrix theories, we use notation $(G_aG_b)G_c$, where $a$, $b$ and $c$ refer to subscripts (or lack thereof) in (\ref{EQ:sigma},\ref{EQ:chi0}).
These theories can be categorized in several different ways. The simplest classification is in terms of 
theories that are closed with a bare propagator line, in contrast to those that are closed with a renormalized
propagator. Sometimes, this simple grouping is adequate to make progress in understanding the success
of various properties of the normal state \cite{beachpla}.
However, historically the other taxonomies have appeared in a systematic way.
The simplest theory is the non-self-consistent theory, that is $(G_0G_0)G_0$.
It has been proposed by Thouless \cite{thouless:1960} to elucidate the nature of superconducting instability that arises in the normal state when approaching the superconducting transition temperature from above.
Unfortunately, a number of difficulties have been reported regarding this approximation \cite{schmitt, serene, sofo}.
A different theory is found when interactions are included via the equation of motion method, and are
represented by $(GG)G_0$ --
this prescription for calculating the single electron self-energy has been discussed by Kadanoff-Martin \cite{kadanoff:1961}. 
Again, some aspects of this approximation have been recognized as not acceptable, and
a discussion of a different approach is contained in the theory introduced by Patton \cite{patton}, that is $(GG_0)G_0$.
The implication of that work is that the ``difficulties" introduced with the Kadanoff-Martin prescription are repaired.

Other theories correspond to self-energies that are closed with a fully renormalized propagator line -- that is $G_c=G$.
One of the important theories that falls into this class is the so-called fully self-consistent T-matrix approximation, $(GG)G$, proposed by Baym \cite{baym:1962} and extended by Bickers \cite{bickers:1,bickers:2}.
Even though this theory might be expected to be the most accurate self energy, simply because it contains the largest number of diagrams, it has been pointed out that it may not be
a wise choice \cite{baym:19xx}.

To decide on which level of self-consistency of the approximate T-matrix theories is required to best reproduce the exact results we need to compare several thermodynamic quantities that can be derived from the modified Hubbard Hamiltonian of (\ref{eq:ZlHam}).
One of these quantities is the expectation value of the electron density per site, which can be determined from
\begin{equation} \label{EQ:density}
n=\frac{2T}{L} \sum_{\vec{k}, m} G(\vec{k}, i \omega_m) \exp(i \omega_m 0^+).
\end{equation} 
The above equation includes the relationship between the particle density and the chemical potential.
Equivalently, one can keep the particle density fixed and adjust the chemical potential so that the above equation was fulfilled, and this latter procedure shall be used below.

We also want to examine how well the approximate theories reproduce thermodynamic quantities that take into account two-particle correlations. One way to do this is to calculate the average double occupancy of the system, which can be evaluated from the expression
\begin{equation} \label{EQ:double_occupancy}
\langle n_{\uparrow} n_{\downarrow} \rangle= \frac{n^2}{4}+
\frac{T}{U L} \sum_{\vec{k}, m} \Sigma(\vec{k}, i \omega_m) G(\vec{k}, i \omega_m) \exp(i \omega_m 0^+),
\end{equation}
This quantity is proportional to the system's interaction (potential) energy.

We also examine the total energy of the system, given by
\begin{equation} \label{EQ:energy}
E=\frac{2T}{L} \sum_{\vec{k}, m} \left[\epsilon_{\vec{k}} +\frac12 \Sigma(\vec{k}, i \omega_m)\right] G(\vec{k}, i \omega_m) \exp(i \omega_m 0^+).
\end{equation}

\section{ \label{SEC:RHM} Repulsive Hubbard Model Results }

\subsection{Unphysical transition of the modified Hubbard Hamiltonian}
 
At first glance, studying the RHM, in comparisons such as those here and in an earlier paper \cite{verga},
seems to be easier than similar analysis for the AHM \emph{when} one is well away from 1/2 filling.
The Thouless criterion does not apply, as the divergence in the pair susceptibility cannot be developed -- its denominator is always non-zero.
On the other hand, the RHM is unstable towards creation of antiferromagnetism at half-filling,
but the antiferromagnetic order is expected to be destroyed very fast when the hole concentration is increased. Other
transitions are also possible \cite{hlubina}, as discussed in the introduction.
However, we have found that the usage of the modified Hamiltonian of (\ref{eq:ZlHam}) induces a phase transition of a different kind.
The reader has to be advised that this kind of behaviour is not present in the (non-modified) Hubbard model of 
(\ref{EQ:hamiltonian_Hubbard})
and is simply an artifact of the procedure required to obtain the results for the T-matrix theories.
Therefore, this artificial phase transition should not be associated with any physical system.
Below we describe the origin of this behaviour, since it is important in the comparisons that we make between
exact and T-matrix theories.

It is straightforward to give analytical expressions for the exact solution of the thermodynamic quantities for the Hubbard Hamiltonian given by (\ref{EQ:hamiltonian_Hubbard}),
or for the modified Hubbard Hamiltonian in (\ref{eq:ZlHam}), when one works in the atomic limit.
(By the atomic limit we understand the reduction of the problem to a single-site by taking the hopping between sites to be zero: $t/U \to 0$).
Note that one may also interpret this limit as being that for which the strength of the Hubbard interaction is infinite,
and one is in the infinitely strong coupling regime -- as we see below, in both this section and the subsequent work described for the AHM, this view of the single-site problem will be important. 

There are only $4$ possible states that can be realized in a single site system, and the grand partition function of the interacting system,
described by the conventional (non-modified) Hubbard Hamiltonian of (\ref{EQ:hamiltonian_Hubbard}), is found easily and yields
\begin{equation}
\mathcal{Z}(T,\mu) = 1 + 2 \exp\left( \frac{\mu}{T} \right) + \exp\left( \frac{2\mu-U}{T} \right).
\end{equation}
The thermodynamic potential of the interacting system can be derived from standard thermodynamic relation $\Omega(T,\mu) = -T \ln \mathcal{Z}(T,\mu)$,
which results in the average particle density given by
\begin{equation}
n \equiv \langle{\hat n}\rangle(T,\mu) = -\frac{d\Omega(T,\mu)}{d\mu} = \frac{1}{\mathcal{Z}(T,\mu)} \left\{ 2 \exp\left(\frac{\mu}{T}\right) + 2 \exp\left(\frac{2\mu-U}{T}\right)\right\}.
\end{equation}
For the purpose of comparing the exact results with various taxonomies of the T-matrix approximation we are interested in the constant density contours of the chemical potential.
Therefore, we invert the above relation to get the chemical potential in terms of temperature and particle density
\begin{equation}
\mu(T,n) = T \ln \frac{1-n + \sqrt{(n-1)^2 + n(2-n) e^{-U/T}}}{(n-2) e^{-U/T}}.
\end{equation}

\begin{figure}[ht!]
\includegraphics[scale=1.1]{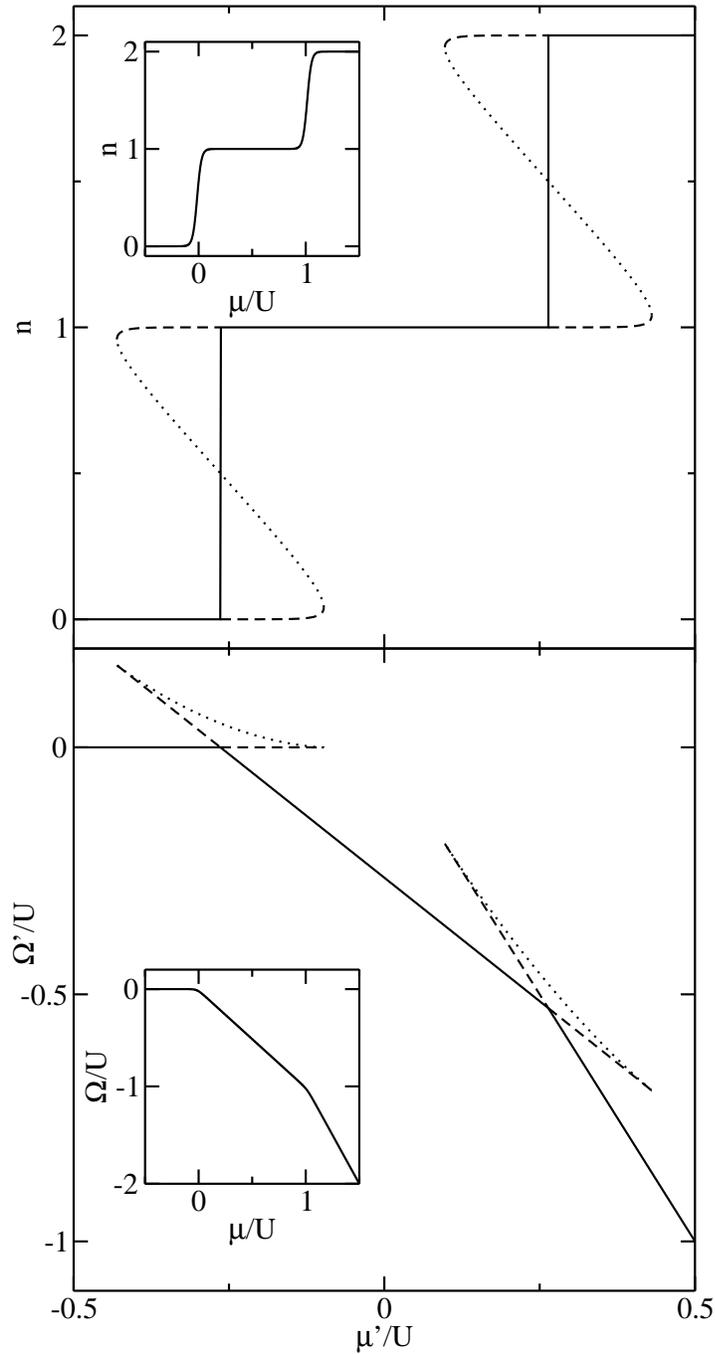}
\caption{ \label{FIG:zlatic}
Exact solution of the RHM in the atomic limit for particle densities $n$ (upper figure), and the corresponding thermodynamic potential 
$\Omega'$  (lower figure), in terms of chemical potential $\mu'$ for the fixed temperature of $T=0.02U$. The thermodynamic
potential and modified chemical potential are both given relative to $U$.
In both figures solid lines corresponds to the stable phase, dashed lines to the metastable phase, and dotted lines to 
the unstable phases. 
The inserts show the exact solution for particle density and thermodynamic potential for the Hubbard Hamiltonian in the atomic limit written in a conventional way,
at the same temperature of $T=0.02U$. }
\end{figure}

The grand partition function can be also easily obtained from the modified Hubbard Hamiltonian of (\ref{eq:ZlHam})
\begin{equation}\label{eq:Zp}
\mathcal{Z}'(T,\mu') = \mathcal{A}(T,\mu') \exp\left(-\frac{Un^2}{4T}\right),
\end{equation}
where
\begin{equation}
\mathcal{A}(T,\mu') = 1 + 2\exp\left(\frac{\mu'+Un/2}{T}\right) + \exp\left(\frac{2\mu'+U(n-1)}{T}\right).
\end{equation}
The thermodynamic potential of the interacting system is then given by
\begin{equation}
\Omega'(T,\mu') = U\frac{n^2}{4} - T \ln \mathcal{A}(T,\mu').
\end{equation}
Note that the second term in the above expression is equal to the thermodynamic potential of a system described by the (non-modified) Hubbard Hamiltonian,
that is by (\ref{EQ:hamiltonian_Hubbard}), if one replace $\mu'$ with $\mu-Un/2$.
This means that the Zlati\'{c} \emph{et al.} transformation \cite{zlatic} is, in fact, transforming the thermodynamic potential into new coordinates.
That is, the thermodynamic potential for the conventional Hubbard Hamiltonian, $\Omega(T,\mu)$, is being transformed into
\begin{equation}
\Omega'(T,\mu') = U \frac{n^2}{4} + \Omega(T,\mu').
\end{equation}
Then, as was done in \cite{verga}, one can find the average particle density of the system from the standard thermodynamic relation $n=-\partial\Omega'(T,\mu')/\partial\mu'$.
Obtaining the chemical potential in terms of particle density then yields
\begin{equation}\label{eq:mp1}
\mu'(T,n) = \frac{U}{2}(1-n)-T\ln \frac{(1-n) e^{U/2T} + \sqrt{n(2-n)+(1-n)^2 e^{U/T}}}{n}.
\end{equation}
However, one can't write an analytical expression for the particle density as a function of chemical potential, $\mu'$, simply by inverting the above relation. 
This is due to the unphysical phase transition that is found at low temperatures, specifically the appearance of van der Waals loops, and which we now describe.

In figure \ref{FIG:zlatic} we show an isotherm for the electron density and the corresponding thermodynamic potential \emph{vs.} chemical potential at $T=0.02U$.
To illustrate how different the results are for the two hamiltonians given by (\ref{EQ:hamiltonian_Hubbard},\ref{eq:ZlHam}), we show the same set of data for the conventional Hubbard model in the insets in this figure.  Clearly,
for the usual Hubbard model (for small finite chains) nothing unusual occurs; only for the modified Hamiltonian does
one obtain something reminiscent of a first-order phase transition. Specifically, we have found that below a critical temperature, the isothermal $n(\mu')$ curve develops a loop, similar
to that observed in the van der Waals equation of state of a weakly interacting gas. 
The thermodynamic potential shows two cusps as a function of chemical potential, $\mu'$, below the transition temperature, and these cusps correspond to inflection points on the particle density curve.
As one can see in the particle density figure, between the inflection points the compressibility is negative,
implying a thermodynamically unstable situation. That can be explained analytically using the transformation
given in (\ref{eq:ZlaticTran}) to express the compressibility \emph{vs.} $\mu'$ in terms of
compressibility in terms of $\mu$. One finds
\begin{equation}
\frac{\partial n}{\partial \mu'}=\frac{\displaystyle \frac{\partial n}{\partial \mu}}{\displaystyle 1-\frac{U}{2} \frac{\partial n}{\partial \mu}}.
\end{equation}
It is then easy to see that when the compressibility of the system described by the conventional Hubbard Hamiltonian satisfies 
$\partial n/\partial \mu>2/U$, the compressibility of the modified Hubbard model becomes negative. Similar behaviour is observed (not shown) 
for chains of all lengths that we studied.
This instability is intrinsic to the RHM when the modified Hamiltonian given by (\ref{eq:ZlHam}) is studied,
but since it is not present in the original Hubbard model  (\ref{EQ:hamiltonian_Hubbard}) it should not be associated with any physical system.

It is important to understand that this behaviour reflects the exact solution for the modified Hamiltonian examined in
the grand canonical ensemble with $\mu'$ as a control parameter, and we are required to use this form of the Hamiltonian
if we are to avoid the issues associated with the Hartree diagrams discussed in this previous sections. Therefore,
this ``phase transition" is used to assess the accuracy of the various T-matrix variants, but to be very clear it is not representative of the physical behaviour of this system, the latter of which could be examined using (1,2). 

\subsection{Comparisons of T-Matrix Results to Exact Thermodynamic Quantities}

\begin{figure}[ht!]
\includegraphics[scale=1.1]{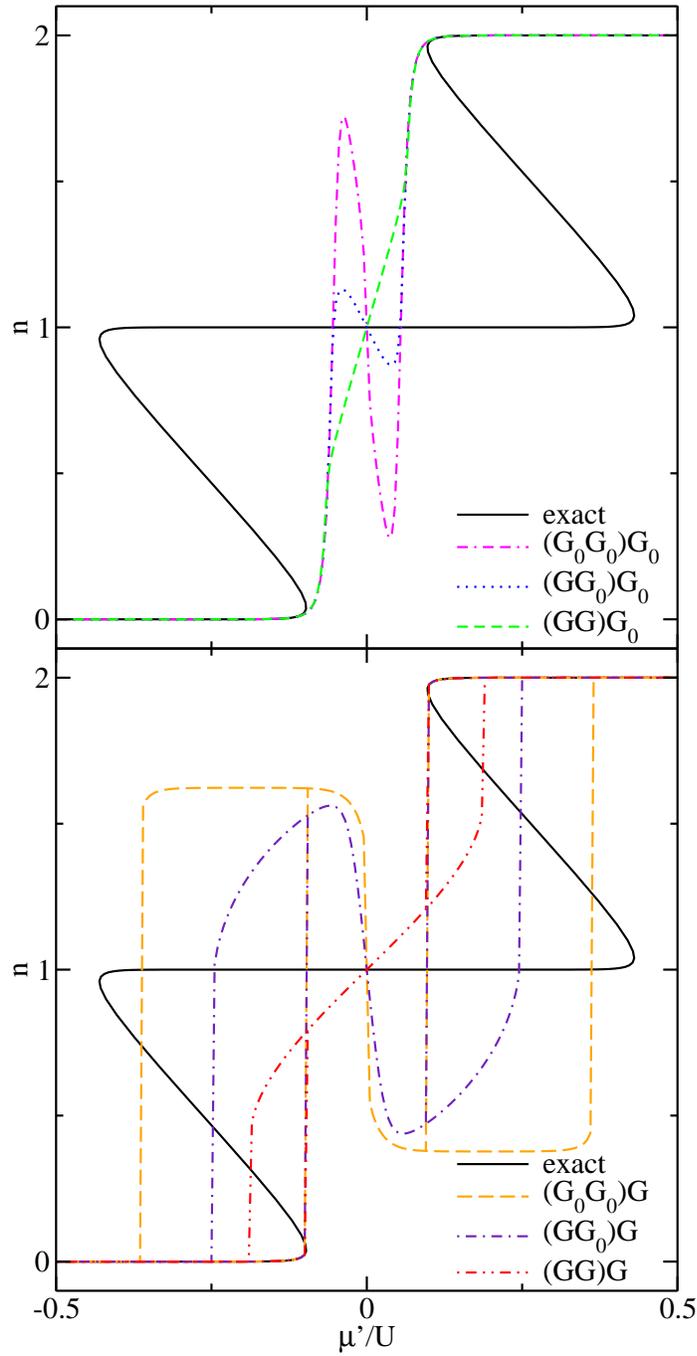}
\caption{ \label{FIG:RHM_mu}
A comparison of the particle density $n(\mu')$ obtained from the exact solution of the RHM with various taxonomies of the 
T-matrix approximation in the atomic limit. The temperature is fixed at $T=0.02U$.
The top figure compares taxonomies corresponding to self energies closed with the non-interacting Green's function, $G_0$, and the bottom figure to taxonomies with self energies closed with a fully dressed propagator line, $G$. As is seen from
a comparison of the two figures, only when the self energy is closed with a fully renormalized $G$ does one obtain
the hysteretic loops found in the exact solution -- see the discussion of the previous figure. }
\end{figure}

\begin{figure*}[p!]
\includegraphics[scale=0.85]{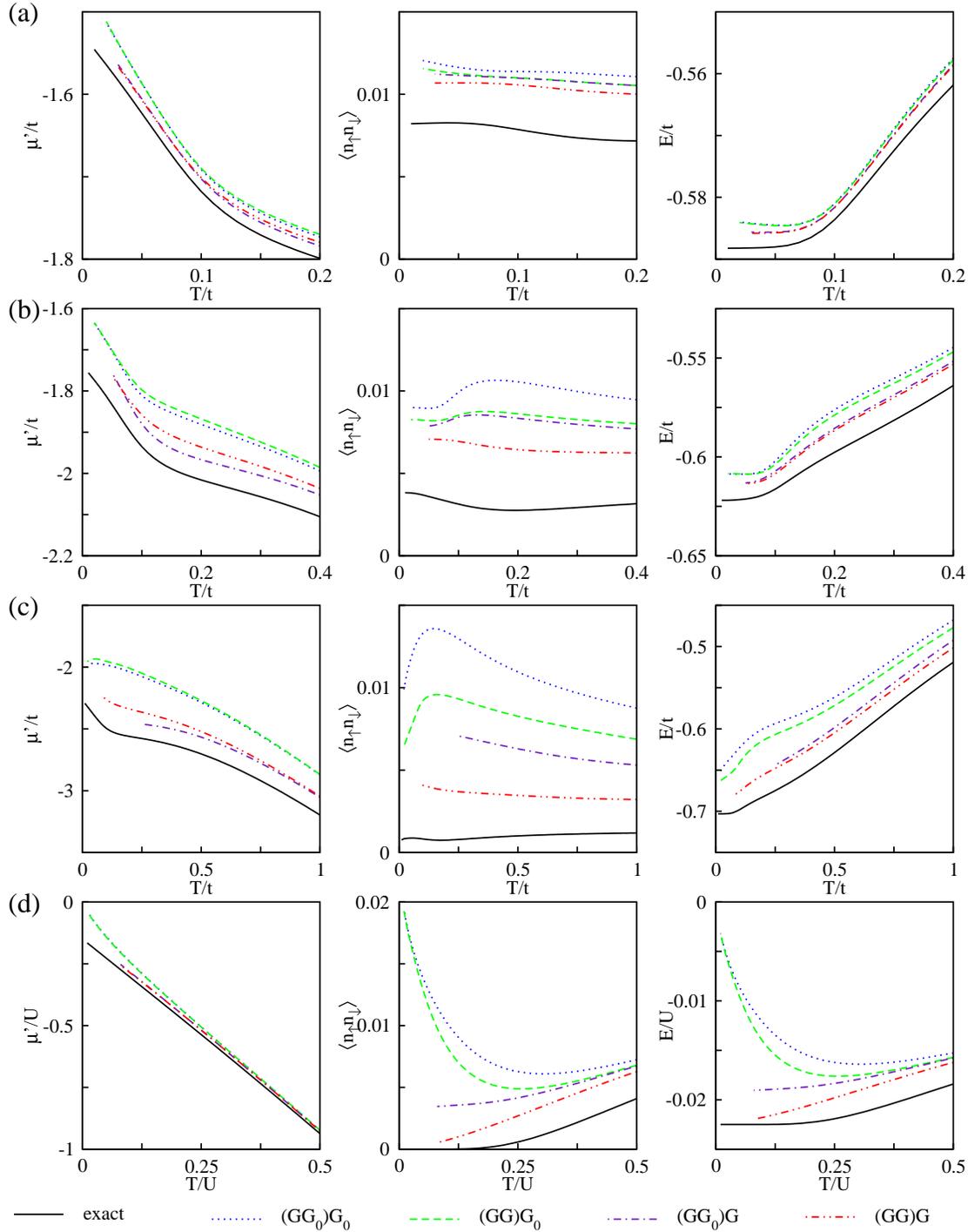}
\caption{ \label{FIG:RHM_T_8sites}
The exact solution of the RHM at a fixed particle density (per lattice site) of $n=0.3$, compared with various taxonomies 
of the T-matrix approximation for a chain of length $L=8$ sites.
These figures show, respectively, the modified chemical potential, $\mu'$, average double occupancy per single site, $\langle n_{\uparrow} n_{\downarrow} \rangle$, and the energy of the system, $E$, for Hubbard interaction strengths of (a) $U/t=2$, (b) $U/t=4$, (c) $U/t=8$ and (d) the atomic limit, which corresponds to $t/U\to 0$.
The various T-matrix taxonomies are labeled according to (\ref{EQ:sigma},\ref{EQ:chi0}). }
\end{figure*}

Now we present our comparisons of the exact result with various T-matrix theories.
As explained earlier, the conventional form of the Hubbard model leads to a great number of Hartree diagrams that must be included in the analysis, and which are troublesome.
Therefore, as in \cite{verga} to complete this comparison we choose to work with the modified Hubbard Hamiltonian of 
 (\ref{eq:ZlHam}).
In figure \ref{FIG:RHM_mu} we show the comparison at $T=0.02U$ for the atomic limit.
We find that the unphysical van der Waals loops of the exactly diagonalized modified Hubbard Hamiltonian of (\ref{eq:ZlHam}) are also reproduced by the class of the taxonomies
for which the self-energy is closed with the fully renormalized Green's function, that is, in (\ref{EQ:sigma}) $G_c=G$.
Therefore, performing the calculation in this manner, one cannot access particle densities that
would result from the usual Hubbard Hamiltonian, since some densities would correspond to unstable phases of 
the above-mentioned loops.
To be concrete, one could use the transformation (\ref{eq:ZlaticTran}) and plot the results in terms of $\mu$, but
this would still result in omissions
(the absence of fully self-consistent solutions at very low temperatures -- see Fig.~\ref{FIG:RHM_T_8sites}) in the curves.
Similar behaviour is observed for all lengths of the chains which we studied -- this is not an artifact of
working in the atomic limit, as we discuss below.

We now present our results for linear chain systems of length $L$ with periodic boundary conditions.
One set of comparisons of the various T-matrix theories with exact solutions that we have completed was for fixed particle density.
We have chosen three values of the Hubbard interaction strength, namely $U/t=2$, $4$, $8$, which correspond to weak, intermediate and strong coupling, respectively.
To find the exact solution of the model Hamiltonian in (\ref{eq:ZlHam}) for a system of multiple sites we use the software provided by the ALPS project \cite{ALPS} and worked in the grand canonical ensemble.
This software allows us to analyze various lengths of the chains.
Of course, the exponential growth of the Hilbert space with the number of particles and the lattice size leads to computational limitations, and we have analyzed system sizes ranging from $L=2$ up to $8$ sites.

Figure \ref{FIG:RHM_T_8sites} shows the comparisons of constant density contours of the modified chemical potential, $\mu'$, the average double occupancy per single lattice site, $\langle n_{\uparrow} n_{\downarrow} \rangle$, and the
energy of the system, $E$, as a function of temperature $T$ for a linear chain of length 8. In all of these plots,
the electron density is fixed at $n=0.3$ (recalling that we are working in the grand canonical ensemble),
again as in \cite{verga}. Because of the behaviour of the isothermal particle density curve that we have just described in the previous subsection, we cannot access all the band fillings below the critical point. To better
facilitate comparisons and follow our conclusions, in figure \ref{FIG:RHM_T_8sites}(d) we also show results for the atomic limit. 

The theory that emerges as being  particularly accurate for the case of the RHM is the fully self-consistent one, $(GG)G$.
Note that the chemical potential obtained within this theory follows closely the exact result in a very wide temperature range (although the $(GG_0)G$ theory is also ``competitive" for the chemical potential).
This agreement is also found for other strengths of the interaction, from weak to strong coupling for linear chain systems.
Proceeding further, we have also examined the thermodynamic properties that require consideration of two particle correlations.
Beginning with the atomic limit and progressing through all strengths of the interaction (or coupling), the fully self-consistent theory again emerges as the most accurate.
It gives the best approximation of the exact solution of the average double occupancy and system's energy, even for the intermediate and strong coupling regimes, which is outside of the domain in which one would expect the T-matrix theory to be
close to the exact results. Note that the \emph{qualitative} behaviour of the exact solution is always reproduced by
the fully renormalized theory, and these results are always the most accurate among all of the analyzed theories.

We arrive at the conclusion that the naive expectation, namely that including more diagrams in Dyson's equation should result in a more accurate approximation, appears to be true for the case of the RHM. The fully self-consistent T-matrix theory most precisely tracks the analyzed thermodynamic quantities: chemical potential, average double occupancy and energy of the system. 

\section{ \label{SEC:AHM} Attractive Hubbard Model Results}

The second case for which we compared the exact solution of the Hubbard model for small linear chains with those
generated by the various T-matrix approximations was for the AHM. As we have mentioned before, the exact solution of the AHM in the atomic limit and the comparison with the various taxonomies of the T-matrix approximation has already been discussed by one of us and colleagues in an earlier paper \cite{verga},
where it was shown that a minimally self-consistent theory, namely the $\Sigma \sim (GG_0)G_0$ T-matrix theory, is particularly accurate.
Here we want to extend those studies for other coupling strengths and to lattices beyond the atomic limit, and to thereby investigate if this minimally self-consistent theory is still the best approximation for the AHM. As we will show, we find that this is only true in the strong-coupling limit, namely the limit in which one is working when examining the model in the atomic limit.

\begin{figure*}[p!]
\includegraphics[scale=0.85]{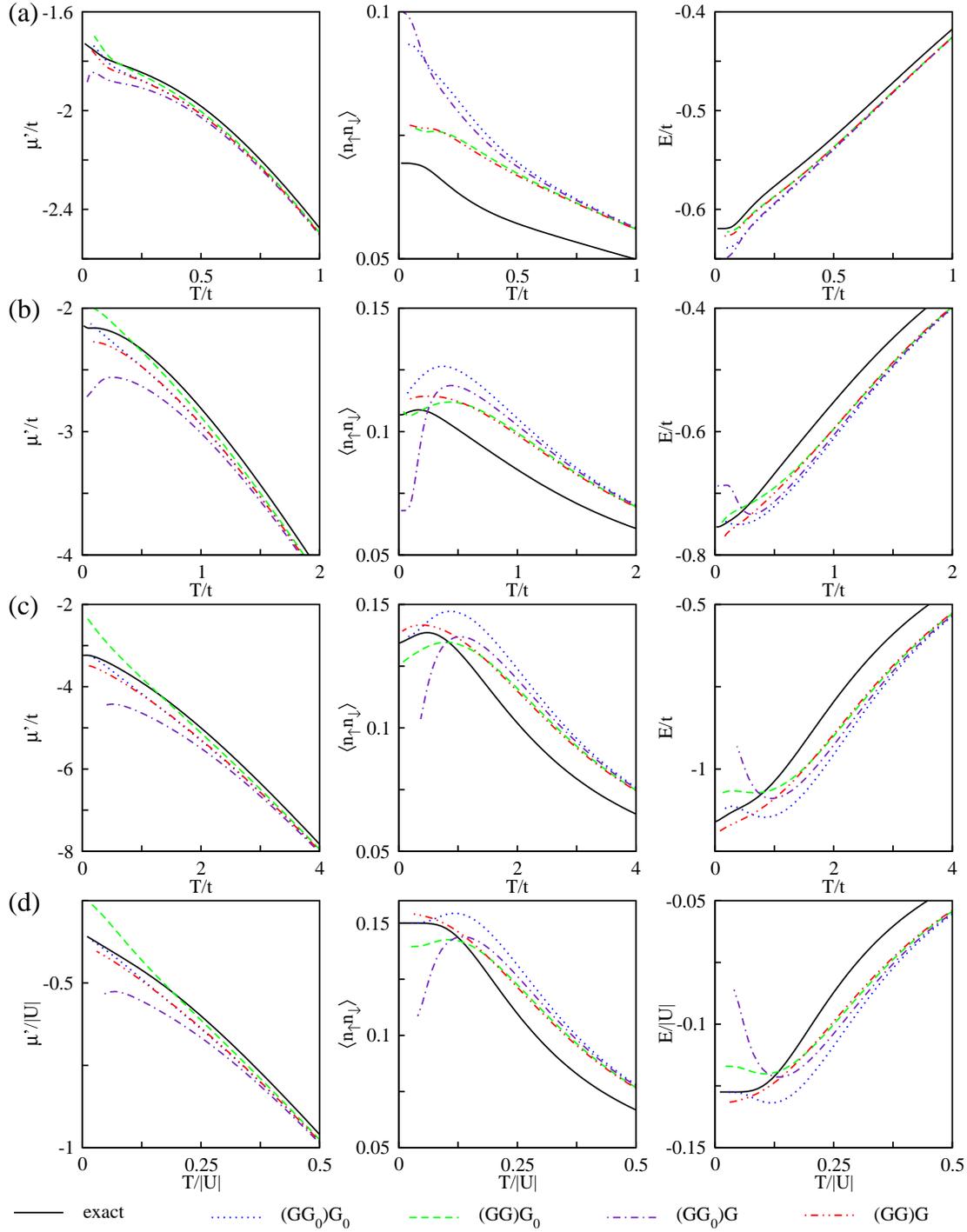}
\caption{ \label{FIG:AHM_T_8sites}
The exact solution of the AHM at fixed particle density per single lattice site of $n=0.3$, compared with various taxonomies of the T-matrix approximation for a chain lattice of $L=8$ sites.
The figures show the modified chemical potential, $\mu'$, average double occupancy per single site, $\langle n_{\uparrow} n_{\downarrow} \rangle$, and energy of the system, $E$, respectively, for Hubbard interaction strength of (a) $U/t=-2$, 
(b) $U/t=-4$, (c) $U/t=-8$ and (d) the atomic limit $t/U\to 0$.
The various T-matrix taxonomies are labeled according to (\ref{EQ:sigma},\ref{EQ:chi0}). }
\end{figure*}

We have diagonalized the Hamiltonian matrix for the atomic limit, and for one-dimensional chains of length $L$ with periodic boundary conditions, which allows us to investigate various strengths of the Hubbard interaction. As for the case of the RHM, we have examined system sizes up to $L=8$ sites and for the same set of (absolute) values of interaction strengths as in the previous section: $|U|/t = 2,~4,~8$. In figure \ref{FIG:AHM_T_8sites} we show the comparisons of the constant density contours for the modified chemical potential, $\mu'$, average double occupancy per single lattice site, $\langle n_{\uparrow} n_{\downarrow} \rangle$, and the energy of the system, $E$, as a function of temperature $T$. (The atomic limit is shown for comparison only -- see below -- the results are equivalent to
those of \cite{verga}.) For very low electron densities (not shown)  we find that all versions of the T-matrix theories work reasonably well. To show how well a particular theory compares we obtained data for a fixed particle density of $n=0.3$ per single lattice site (as in \cite{verga}), which we take as a density at which one would hope that the T-matrix approximation would be reasonably accurate.

The success of the minimally self-consistent $(GG_0)G_0$ theory in the atomic limit is seen in the lowest row of data. Specifically, for $T \lesssim 0.05 |U|$ this theory quickly converges to the exact result  for all three quantities shown. In the
data shown (as well as for the other lattice sizes that we have studied, which are not shown) in the strong coupling case
of $|U|/t=8$ (second bottom row of data) we find similar agreement at low temperatures. This is not unexpected,
since the $|U|/t \rightarrow \infty$ limit yields uncoupled sites, namely the atomic limit. As these two sets of data
make clear, only at the lowest temperatures is this agreement found. At low temperatures the two taxonomies that
correspond to a self energy closed with a fully renormalized propagator, namely $\Sigma \sim (GG_0)G$  and 
$\Sigma \sim (GG)G$, are seen to be closer to the exact result, and, in fact, the minimally self-consistent theory performs the least successfully. As we discuss below, this may be important in the pair fluctuation regime of the anomalous normal phase.

The top two rows of data show our results for weak and intermediate coupling. We find that while the $(GG_0)G_0$ theory
is still quite good at representing the variation of $\mu'$ at low temperatures, the two-particle correlations, and also
the energy, are certainly not most accurately represented by this theory. One would expect a diagrammatic expansion
such as the T-matrix theory to be most successful for weak interactions. As our data indicates, and as most clearly shown
by the site-averaged $\langle n_{\uparrow} n_{\downarrow} \rangle$, for weak coupling (top row of data) the $(GG_0)G$ and $(GG)G$ theories are much closer to the exact data for temperatures less  than 0.5$t$. By examining all three thermodynamic
quantities, in fact, one would say that like the RHM the fully renormalized theory is the most successful  
in the limit of weak coupling! Further, this theory is most successful in the full temperature range, not simply the
low temperature agreement found for the $(GG_0)G_0$ theory for the strong coupling and atomic limits.
(We again emphasize that the $(G_0G_0)G_0$ and $(G_0G_0)G$ data are not shown, but are the least
successful variants of T-matrix theory in reproducing the exact data, the same as for the RHM.)

\begin{figure}[ht!]
\includegraphics{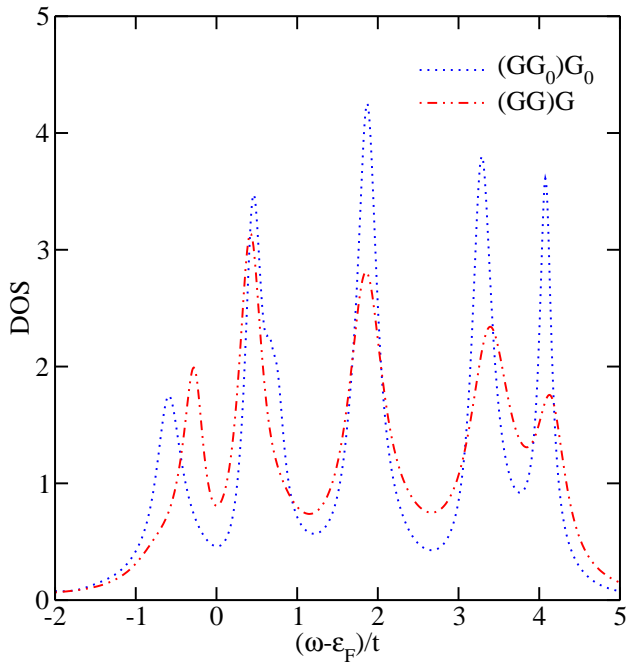}
\caption{ \label{FIG:AHM_DOS}
The density of states (DOS) evaluated in the $(GG_0)G_0$ and $(GG)G$ T-matrix approximations for a chain lattice of $L=8$ sites and a fixed particle density per single site of $n=0.3$.
The temperature is set at $T=0.1t$ and the strength of Hubbard interaction at $U/t=-2$, which corresponds to weak coupling.
The peaks of the spectral functions, used to obtain the density of states, have been broadened using a Lorentzian function with the half-width parameter fixed at $\eta=0.06t$. }
\end{figure}

To investigate further the candidate taxonomies that emerge as the most accurate, that is $(GG_0)G_0$ and $(GG)G$, we have evaluated the density of states of the system. Figure \ref{FIG:AHM_DOS} shows the density of states (DOS), which we have
determined via
\begin{equation}
\label{EQ:DOS}
DOS(\omega)=-\frac{1}{\pi} \sum_{\vec{k}} \mathrm{Im} A(\vec{k},\omega),
\end{equation}
where the spectral functions $A(\vec{k},\omega)$ for each wave vector have been obtained through 
a rational Pad\'{e} approximant \cite{beach-pade}. 
Our results point out that the peaks in the DOS obtained for the minimally self-consistent $(GG_0)G_0$ theory have a minimum at the Fermi level that is deeper than for the case of a fully renormalized $(GG)G$ theory.
Therefore, a pseudo-gap type of depression of the DOS is found to be much more prominent in the minimally self-consistent $(GG_0)G_0$ theory than in a fully renormalized variant of the T-matrix approximation. 

\section{ \label{SEC:conclusions} Discussion }

We have analyzed various T-matrix theories of a modified Hubbard model in the dilute limit, and compared them with the exact solution of both the AHM and RHM. The analyzed cases concern the atomic limit, as well as finite one-dimensional lattices for which we have analyzed a variety of the coupling strengths:  weak, intermediate and strong.

We have shown that the modified Hubbard model gives rise to an unphysical first-order phase transition when working with the RHM. This behaviour serves as a diagnostic when comparing the various T-matrix variants.The hysteretic behaviour that we find in the exact solutions appears in versions of the T-matrix approximation that close the self energy with a fully renormalized single-particle propagator, both in the atomic limit, and
for finite lattices for all coupling strengths. In contrast to this, self energies that are closed with a bare
propagator do not find the instability that is present only in this modified model Hamiltonian.
Our comparisons of other thermodynamic quantities examined for the case of the RHM make evident that the fully renormalized theory,  $(GG)G$, is the most successful T-matrix theory for this model at moderately low electronic densities. That is, this theory reproduces the data obtained from the exact diagonalization of the Hamiltonian 
matrix for a wide temperature range, and also for wide variety of coupling strengths.   

For the case of AHM we find that the success of the minimally self-consistent T-matrix theory, $(GG_0)G_0$, found in
the atomic limit \cite{verga} in the regime of low temperatures is not reproduced for finite lattices at weaker coupling. In
fact, our results support that for weak coupling fully renormalized theories are better at all temperatures studied
($T \lesssim 0.5 t$). Further, the density of states of minimally \emph{vs.}  fully renormalized theories is found to be different,
in that the former has a larger depression of the DOS at the Fermi energy.
Since one expects T-matrix diagrammatic theories to be most accurate in the low coupling regime, this result casts some doubt on the applicability of approaches based on
a  $(GG_0)G_0$ variant of the T-matrix approximation for the AHM.
However, our results are for very small one dimensional lattices (where exact comparisons are possible),
and further work with larger lattices in higher dimensions is required before firm conclusions are possible.

Recently a new variation of a T-matrix type of theory was proposed by \v{S}op\'ik \emph{et al.} \cite{sopik} This new formulation is based on a T-matrix approximation in which one removes certain ``dangerous" contributions to the self energy arising from a resonance of an interacting pair (see their paper for further details). It will be interesting to see if ongoing numerical comparisons \cite{privcomm} of this new theory to the exact
results (for the AHM) lead to improvements in reproducing thermodynamic and dynamical properties.

\ack
We acknowledge helpful conversations with Ryan Jones regarding the implementation of the ALPS software package. 
This work was supported in part by the NSERC of Canada.

\newpage

\section*{References}

\bibliographystyle{jpcm}
\bibliography{bibliography}{}

\end{document}